\documentclass[aps,prl,twocolumn,groupedaddress,showpacs]{revtex4}
\usepackage{graphicx}
\usepackage{float}
\usepackage{dcolumn}
\usepackage{amsmath}
\usepackage{amssymb}
\input{epsf}

\newtheorem{theorem}{Theorem}

\newtheorem{claim}[theorem]{Claim}

\newcommand{\qedsymb}{\hfill{\rule{2mm}{2mm}}}
\newenvironment{proof}[1][]{\begin{trivlist}
\item[\hspace{\labelsep}{\bf\noindent Proof#1:\/}] }{\qedsymb\end{trivlist}}

\newcommand{\ket}[1]{\left| {#1} \right\rangle}
\newcommand{\bra}[1]{\left\langle {#1}\right |}
\newcommand{\iprod}[2]{\left\langle {#1}|{#2}\right\rangle}

\newcommand\abs[1]{{\left| {#1} \right|}}

\def\be{\begin{equation}}
\def\ee{\end{equation}}
\def\bea{\begin{eqnarray}}
\def\eea{\end{eqnarray}}

\def\>{\rangle}
\def\<{\langle}

\begin{document}
\title{Adiabatic Rotation, Quantum Search and Preparation of
Superposition States\footnote{This work was supported by the
U.~S.~DOE, Contract No.~DE-AC02-76SF00515.}}
\author{M. Stewart Siu\footnote{msiu@stanford.edu}}
\address{Stanford Linear Accelerator Center, Stanford University,
  Stanford, California 94309}
\date{\today}

\begin{abstract}
    We introduce the idea of using adiabatic rotation to generate
    superpositions of a large class of quantum states. For quantum
    computing this is an interesting alternative to the well-studied
    "straight line" adiabatic evolution. In ways that complement recent
    results, we show how to efficiently prepare three types of states:
    Kitaev's toric code state, the cluster state of the
    measurement-based computation model and the history state used in
    the adiabatic simulation of quantum circuit. We also show that the
    method, when adapted for quantum search, provides quadratic speedup
    as other optimal methods do with the advantages that the problem
    Hamiltonian is time-independent and that the energy gap above the
    ground state is strictly nondecreasing with time. Likewise the
    method can be used for optimization as an alternative to the
    algorithm of Farhi et al \cite{fggs}.
\end{abstract}

 \maketitle

\section{I. Introduction}

The presence of superpositions of classical states lies at the heart
of many non-classical behaviors of quantum systems. It is therefore
not surprising that uniform superposition states serve as powerful
resources for quantum information processing. The main result in
this article is that there is a conceptually simple way of preparing
a class of linear superpositions by adiabatically rotating a driving
Hamiltonian. Seen another way, it is also a method for finding the
ground states of certain Hamiltonians. While the applications we
discuss are all related to quantum computing, the idea is relevant
to preparation of quantum states in general.

Adiabatic evolution studied in quantum computing often takes the
form of a linear interpolation $H(s)=(1-s)H_{initial}+sH_{final}$
whether it is for problem solving \cite{fggs,rc} or state
preparation \cite{lidar,BR,adkllr}. In this article we consider a
different paradigm that is inspired by observations in \cite{c2a}
and follows a "rotation" instead. A time-dependent similarity
transform on the entire Hamiltonian, as pointed out in \cite{c2a},
often requires highly nonlocal interactions and thus seems to have
few applications. One may however adopt a different perspective and
consider this time-dependent transformation as a driving term in the
presence of a well-gapped and time-independent Hamiltonian. What we
will show in Section II is that some of the desirable properties
associated with the similarity transform remains in this more
general setting.

In Section III, we outline how to use this idea to prepare states
with interesting physical properties. The first example is the
ground states of Kitaev's toric code \cite{kitaev}, which have been
of much interest because of their natural fault tolerance as memory
and exhibition of topological order\cite{wen}. In \cite{lidar} it is
shown that a ground state of the toric code Hamiltonian can be
prepared adiabatically through an linear interpolation in time
$O(\sqrt{n})$ where n is the number of sites. Both this method and
the original preparation-by-measurement method \cite{dennis} are
optimal since they saturate the Lieb-Robinson bound\cite{Hastings},
which places a theoretical limit on the efficiency of any
preparation method for topologically ordered states. However it
requires duality mapping to an Ising model, which means the same
result cannot be easily reproduced for a different system with
topological order. We will show, without requiring extra types of
interactions, that there is an elementary, circuit-like adiabatic
evolution path that also prepares the toric code state optimally.
The same idea would not only be applicable to other topologically
ordered states, but also, for instance the cluster state
\cite{cluster} used in measurement-based computing. Recently it was
shown in \cite{BR} that the a state similar to the cluster state in
computation power can be prepared efficiently via adiabatic linear
interpolation. Our method shows how the cluster state itself can be
prepared adiabatically and efficiently.

In Section IV, we move on to a more complicated example. In
\cite{c2a} it is noted that the adiabatic linear interpolation
corresponding to a quantum circuit used by \cite{adkllr} can be
thought of as a special case in a larger family of paths. It may
still appear, however, that fairly elaborate gap analysis
\cite{adkllr, ruskai} would be required to check the efficiency of
any path that generates the history state. This turns out to be not
the case. We will describe two paths that generate the history state
efficiently, and apart from being amenable to gap analysis, one only
requires time-dependent interactions on a few qubits.

In Section V, we turn our attention to problem solving. Adiabatic
quantum computing was originally proposed not for state preparation
but to solve optimization (especially NP-complete) problems. While
there is still controversy on the value of this approach, it is
well-known that the the closely related adiabatic quantum search
method \cite{rc} yields an optimal result, in the sense that it is
as efficient as Grover's algorithm. Here we present an adiabatic
rotation version of quantum search. Like the linear interpolation
version, it is also optimal. The difference is that in our case, the
spectral gap between the ground state and the first excited state is
strictly nondecreasing, and this feature can be extended to other
optimization problems. Clearly this would make it easier for us to
determine whether a given problem can be solved efficiently. Our
approach, however, has an additional difficulty compared to linear
interpolation. While exact examples of linear interpolation
algorithms often assume nonlocal interactions (as \cite{rc} does),
one can almost always formulate a linear interpolation algorithm
with local interactions. The situation is different with adiabatic
rotation, because it requires separation of the initial state from
the computational states. We will conclude with a brief discussion
of what this means and some suggestions for future direction.

\section{II. The basic formulation}

In the simplest case, our Hamiltonian consists of a time-independent
piece and a driving term under time-dependent similarity transform:
\begin{equation}
H(\theta)=H_0+U_m(\theta)h_mU(\theta)_m^\dag \label{arot}
\end{equation}
where $\theta$ is a time-dependent rotation parameter and $H_0$ is a
Hamiltonian with two degenerate ground states of zero energy.
Suppose one ground state is $\ket{\psi_a}=\sum_{i=1}^n a_i
\ket{\gamma_i}$ and another is $\ket{\psi_b}=\ket{\gamma_m}$ where
$\{\ket{\gamma_1}...\ket{\gamma_n},\ket{\gamma_m}\}$ spans the
effective Hilbert space of interest. The $\theta$-dependent term
acts on a two-dimensional space spanned by
$\{\ket{\gamma_m},\ket{\gamma_j}\}$ for some $1\leq j\leq n$. That
is, we define $h_m=K\ket{\gamma_m}\bra{\gamma_m}$ for some positive
constant $K$ and $U(\theta)$ is a rotation in
$\{\ket{\gamma_m},\ket{\gamma_j}\}$:
\[
U_m(\theta)=e^{i\textbf{J}\theta}=\left(\begin{array}{cc}
\cos{\theta} & \sin{\theta}\\
-\sin{\theta} & \cos{\theta}\\
\end{array}\right),
\]
\bea U_mh_mU_m^\dag = K\left(\begin{array}{cc}
\cos^2{\theta} & -\sin{\theta}\cos{\theta}\\
-\sin{\theta}\cos{\theta} &\sin^2{\theta}\\
 \end{array}\right) \label{driving}
\eea
 Here $\textbf{J}$ denotes the generator of rotation. One can
check that \be \ket{0}=\frac{1}{\sqrt{1+a_j^2 \tan^2{\theta}}}(a_j
\tan{\theta}\ket{\gamma_m}+\sum_{i=1}^n a_i \ket{\gamma_i}),
\label{ground} \ee is a ground state of $H(\theta)$ with zero energy
for any $\theta$. We will consider $\theta$ between 0 and
$\frac{\pi}{2}$.

In adiabatic evolution, the transition rate from a ground state
$\ket{0}$ to an excited state $\ket{k}$ is well-known to be bounded
by $\abs{\bra{k}\frac{dH}{dt}\ket{0}/(E_k-E_0)^2}^2$\cite{Messiah}.
To keep this quantity constant as we vary $\theta$, we write
$\frac{dH}{dt}=\frac{dH}{d\theta}\frac{d\theta}{dt}$ and it follows
that we need \be \frac{d\theta}{dt}\sim
\frac{(E_k(\theta)-E_0(\theta))^2}{\abs{\bra{k}\frac{dH}{d\theta}\ket{0}}}.\label{speed}\ee
Our model offers at least two advantages if we wish to adiabatically
manipulate a state of the form (\ref{ground}).
\begin{claim}
Let $g_k(\theta)=\abs{E_k(\theta)-E_0(\theta)}$. For
$0<\theta<\frac{\pi}{2}$, $g_k'(\theta)\neq 0$.
\end{claim}
\begin{proof}
If $\ket{k(\theta)}$ is an eigenstate of $H(\theta)$ with energy
$E_k(\theta)$, by first-order perturbation theory, \be
\frac{dE_k(\theta)}{d\theta}\sim
\bra{k(\theta)}\frac{dH(\theta)}{d\theta}\ket{k(\theta)}
\nonumber\ee where $\frac{dH(\theta)}{d\theta}$ can be explicitly
evaluated as
\begin{eqnarray}
\frac{dH(\theta)}{d\theta}&=& e^{i\textbf{J}\theta}[i\textbf{J},H]e^{-i\textbf{J}\theta}\nonumber\\
&=&\left(\begin{array}{cc}
-2\cos{\theta}\sin{\theta} & \sin^2{\theta}-\cos^2{\theta}\\
\sin^2{\theta}-\cos^2{\theta} & 2\cos{\theta}\sin{\theta}\\
 \end{array}\right) \label{dhdt}
\end{eqnarray}
One readily checks that for $0<\theta<\frac{\pi}{2}$ the expectation
value of this operator vanishes if and only if the $\ket{\gamma_m}$
and $\ket{\gamma_j}$ components of the state satisfies the
$\tan{\theta}$ ratio. If an eigenstate of $H(\theta)$ satisfies
this, it would have to be the ground state which stays at zero. This
means the gap only shifts in one direction.
\end{proof}
\begin{claim}
$\abs{\bra{k}\frac{dH}{d\theta}\ket{0}}\leq
O({\sqrt{g_k(\theta)}})$.
\end{claim}
\begin{proof}
The idea of the proof is that $\frac{dH}{d\theta}\ket{0}$ couples to
a state with high energy, and this limits its coupling into low
lying states during the evolution. Using (\ref{dhdt}), we obtain
\bea
\frac{dH}{d\theta}\ket{0}&=&\frac{a_j}{\sqrt{\cos^2{\theta}+a_j^2
\sin^2{\theta}}}\ket{p}, \nonumber \\
\ket{p}&\equiv&\sin{\theta}
\ket{\gamma_j}-\cos{\theta}\ket{\gamma_m}.\nonumber \eea The energy
of an excited state is
$g_k(\theta)=\abs{\iprod{p}{k}}^2K+\bra{k}H_0\ket{k}$, from which we
see that $\abs{\iprod{p}{k}}\leq \sqrt{g_k(\theta)/K}$.
\end{proof}
The first fact enables us to estimate efficiency by looking only at
the end points. The second allows a higher speed for $d\theta/dt$
when the gap is small. A more detailed estimate yields:
\begin{eqnarray}
&&\abs{\bra{k}\frac{dH}{d\theta}\ket{0}}^2\leq
A[(1+r^2)c_{mk}^2g_1-g_1+g_k(\theta)]
\nonumber\\
&&A\equiv\frac{a_j^2}{K(\cos^2{\theta}+a_j^2 \sin^2{\theta})}
\nonumber\\
&&r\equiv\frac{\sum_{i=1}^n
a_i\iprod{\gamma_i}{0}}{\iprod{\gamma_m}{0}},~~
c_{mk}\equiv\iprod{\gamma_m}{k} \label{estimate}
\end{eqnarray}
where $g_1$ is the first excited state energy of $H_0$. We see that
the quantities $A$ and $c_{mk}$ can further suppress the transition
rate as they may contain factors inversely proportional to the size
of the system.

To see why this is useful, consider the case
$a_i=\frac{1}{\sqrt{n}}$ for $i=1..n$ and $\theta$ running from 0 to
$\frac{\pi}{4}$ adiabatically. It is clear that we can turn the
state $\frac{1}{\sqrt{n}}(\ket{\gamma_1}+...+\ket{\gamma_n})$ to
$\frac{1}{\sqrt{n+1}}(\ket{\gamma_1}+...+\ket{\gamma_n}+\ket{\gamma_m})$,
thus adding an element to the uniform superposition. Reverse the
process and we will delete the element from the superposition.
Clearly, repeated applications of such processes with different
driving terms allow us to prepare a large class of states. Generally
speaking, if we wish to create a particular superposition of certain
basis states, we can look at the Hamiltonian interactions we have
control over that act either diagonally or off-diagonally on the
basis. This set of interactions can then be thought of as a
generating set for group elements that we want to eventually appear
in the superposition. In the next section we will consider some
examples to illustrate this idea.

\section{II. The toric code and the cluster state}

\begin{figure}
  \includegraphics[width=3in]{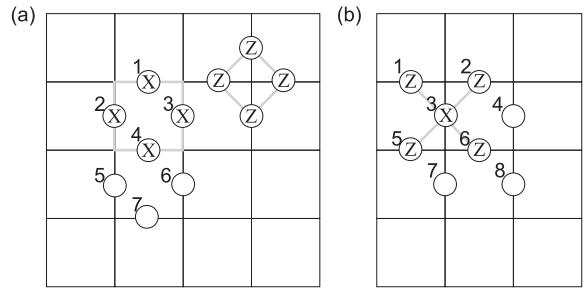}\\
  \caption{(a) Interactions that appear in the toric code
  Hamiltonian. (b) Interactions in the cluster state Hamiltonian,
  following the same notation.}\label{lattice}
\end{figure}

The ground state of the toric code Hamiltonian\cite{kitaev} can be
generated efficiently by adiabatic linear interpolation
\cite{lidar}. The minimal gap in the evolution is estimated to be of
order $O(n^{-\frac{1}{2}})$. Here we show a different path where the
gap remains constant but the evolution time is also $O(\sqrt{n})$,
thus also saturating the Lieb-Robinson bound \cite{Hastings}.
Following the basis choice of \cite{lidar}, the Hamiltonian is: \be
H=~-\sum_{\textrm{~~~star~~}}Z\otimes Z\otimes Z\otimes
Z~~~-\sum_{\textrm{plaquette}}X\otimes X\otimes X\otimes X
\label{toricHam} \ee where $X$ and $Z$ are Pauli matrices and the
spin lattice is as shown in Fig.\ref{lattice}(a). The spins live on
the links (as this is a $Z_2$ lattice gauge theory); a "plaquette"
refers to the four spins around a square while a "star" means the
four spins around a vertex. We will work within the ground states of
the star terms so we can effectively ignore them. Consider the spins
$\{1,2,3,4\}$ in Fig.\ref{lattice}(a) and a $\theta$-dependent
Hamiltonian: \bea
H_p(\theta)&=&1+(\sin^2{\theta}-\cos^2{\theta})(Z_1+Z_2+Z_3+Z_4)/4\nonumber\\
&&-2\sin{\theta}\cos{\theta}X_1X_2X_3X_4 \label{plaq}\eea We can
check that within the subspace of $\ket{0000}$ and $\ket{1111}$ this
has exactly the form of the driving term in (\ref{driving}) and that
it takes $\ket{0000}$ to $\frac{1}{\sqrt{2}}(\ket{0000}+\ket{1111})$
as $\theta$ goes from 0 to $\frac{\pi}{4}$. Other states in the
Hilbert space, regardless of energy, do not affect the result
because they are decoupled from the subspace, so the effective gap
is $2$ throughout the process. Since the ground state of the
Hamiltonian (\ref{toricHam}) is known to be the superposition of all
"closed strings" (plaquette excitations), the goal is to do the same
operation of rotating the $Z$ operators into $XXXX$ for every
plaquette. Consider next the plaquette $\{4,5,6,7\}$ after we have
evolved $\{1,2,3,4\}$. The Hamiltonian we use would have to be a
little different since "$4$" is already entangled: \bea
H_p'(\theta)&=&1+(\sin^2{\theta}-\cos^2{\theta})(Z_5+Z_6+Z_7)/3\nonumber\\
&&-2\sin{\theta}\cos{\theta}X_4X_5X_6X_7 \eea

It is easy to verify that this Hamiltonian takes
$\frac{1}{\sqrt{2}}(\ket{0000}+\ket{1111})\ket{000}$ to
$\frac{1}{2}(\ket{0000000}+\ket{0001111}+\ket{1111000}+\ket{1110111})$
as $\theta$ goes from 0 to $\frac{\pi}{4}$. Generally, we can add a
plaquette excitation to the superposition if at least one out of the
four spins is unentangled. Bearing this in mind, we can perform
these adiabatic rotations on plaquettes in parallel (coefficients in
the Hamiltonian can be also adjusted to give a more translationally
symmetrical appearance). For example, on a square lattice, we can
operate on alternate rows of plaquettes in parallel. Dividing each
row into odd and even plaquettes, this can be accomplished in two
time steps. Then we operate on the remaining rows sequentially - to
make sure we have an unentangled spin for every operation - which
takes $L$ steps if the lattice has $n=L\times L$ spins. The genus of
the surface does not really affect this algorithm, except we have to
notice that because the same loop can be viewed from two sides in a
torus, there will be some plaquette operators that do not lead to
new states in the superposition. As for other topologically
inequivalent ground states, they can be generated in this manner by
reversing the $Z$ operators around an incontractible loop at the
beginning. Compared to the linear interpolation, our evolution path
is certainly more complicated. But it provides an interesting
perspective, as it does not rely on duality mapping to an Ising
model. In principle the technique we use can be generalized to
prepare any string-net condensed state\cite{wen} because they are
uniform superpositions of group elements generated by local
string-nets.

The same trick also works for the cluster state\cite{cluster}.  The
cluster state used in measurement-based quantum computing, if put on
a square lattice, is the superposition
$\frac{1}{2^n}\sum_{z_i=\{0,1\}} (-1)^r\ket{z_1...z_n}$ where $r$ is
the number of adjacent "1"'s in the configuration $\{z_1...z_n\}$.
It is stabilized by the operators $\sum_s X_s (\bigotimes_{s'}
Z_{s'})$ where $s'$ denotes sites adjacent to site $s$ and the
tensor product is over all four adjacent sites. Since each term
takes half the states in the superposition to the other half, they
play a role similar to the plaquette operators in the toric code. We
can set the Hamiltonian for the sites $\{1,2,3,5,6\}$ in
Fig.\ref{lattice}(b) to be something similar to (\ref{plaq}): \bea
H_c(\theta)&=&1+(\sin^2{\theta}-\cos^2{\theta})Z_3\nonumber\\
&&-2\sin{\theta}\cos{\theta}X_3Z_1Z_2Z_5Z_6 \eea and take $\theta$
from 0 to $\frac{\pi}{4}$. The same operation can be repeated on
sites $\{3,4,6,7,8\}$, for instance. But since we always have an
unentangled spin every time we turn on a stabilizer term, unlike the
toric code case there is no need to operate sequentially. Therefore
we can operate on all the odd sites in the first time step and all
the even site in the second time step, completing the algorithm in
$O(1)$ time.

Note that nothing prevents us from applying perturbation gadgets
\cite{gadget} on the stabilizers and turning our Hamiltonian into a
nearest-neighbor one. Thus this is comparable with the linear
interpolation result of \cite{BR}, with the difference that we are
preparing the cluster state rather than a state that can be used as
one.

\section{IV. The history state}

\begin{figure}
  \includegraphics[width=3in]{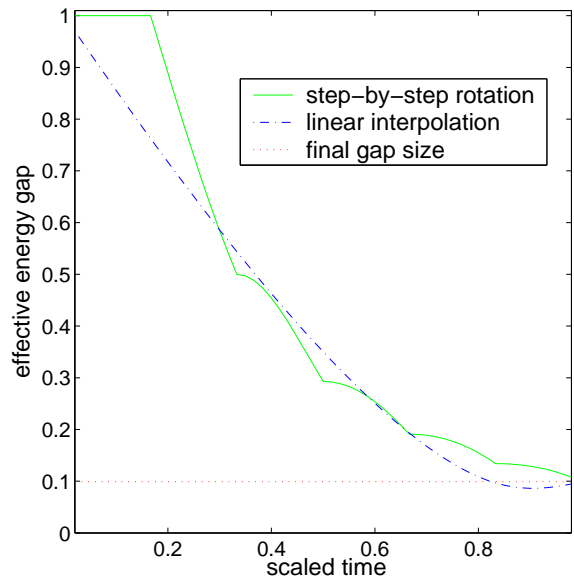}\\
  \caption{Comparison of energy gap between stepwise adiabatic rotation to linear interpolation in the
  generation of history state with L=6. For the stepwise adiabatic rotation, the gap never
  crosses the line corresponding to the final gap. (Color online)}\label{sevenstate}
\end{figure}

In all our examples so far the effective energy gap remains
constant, so essentially we are implementing quantum circuits in the
sense of \cite{c2a}. Now let us consider an example where the gap
shrinks as the system grows. In \cite{adkllr}, it is shown that the
history of a quantum circuit - a uniform superposition of
intermediate states, appropriately made orthogonal - can be
generated via an adiabatic linear interpolation. Up to small
corrections unimportant for our purpose, the initial and final
Hamiltonians are: \bea
H_i&=&\sum_{t=1}^L\ket{\gamma_t}\bra{\gamma_t} \nonumber\\
H_f&=&\frac{1}{2}\sum_{t=0}^{L-1}\ket{\gamma_t}\bra{\gamma_t}+\ket{\gamma_{t+1}}\bra{\gamma_{t+1}}\nonumber\\
&&-\ket{\gamma_t}\bra{\gamma_{t+1}}-\ket{\gamma_{t+1}}\bra{\gamma_t}
\label{histham}\eea where $\ket{\gamma_t}$'s represent the
orthogonal intermediate states at time steps labeled by $t$. This
adiabatic evolution takes the starting state $\ket{\gamma_0}$ to the
history state $\frac{1}{\sqrt{L+1}}\sum_{t=0}^L \ket{\gamma_t}$. An
obvious alternative to linear interpolation from $H_i$ to $H_f$
would be to apply adiabatic rotation step by step and rotate each
term in $H_i$ to a corresponding term in $H_f$: \bea
H_t(\theta_t)&=&\sin^2{\theta_t}\ket{\gamma_t}\bra{\gamma_t}+\cos^2{\theta_t}\ket{\gamma_{t+1}}\bra{\gamma_{t+1}}
\nonumber\\
&&-\sin{\theta_t}\cos{\theta_t}(\ket{\gamma_t}\bra{\gamma_{t+1}}+\ket{\gamma_{t+1}}\bra{\gamma_t})
\eea where as usual, $\theta_t$ goes from $0$ to $\frac{\pi}{4}$ at
each time step. Fig.\ref{sevenstate} shows the energy gap as a
function of time for $L=6$ and compares this stepwise method to the
linear interpolation. We can see that the minimal gap occurs at the
very end. In \cite{adkllr} sophisticated techniques were used to
estimate the scaling of the minimal gap as a function of $L$. It is
much easier in our case, even compared to the simpler analysis of
\cite{ruskai}. Observe that the final Hamiltonian is the discrete
version of a kinetic energy term for a free particle in a box of
size $L$. For large $L$, its spectrum simply consists of standing
waves with energy proportional to the square of momentum, which goes
as the inverse of wavelength. Thus we immediately conclude that the
gap is $O(L^{-2})$. Now consider the speed of evolution as measured
by $d\theta/dt$ for fixed error rate. Naively, it would scale as the
square of the gap, i.e. $O(L^{-4})$. But we can use the bound in
Eq.\ref{estimate} to get a better result. At the minimal gap,
$g_1\sim O(L^{-2})$ and $A\sim O(L^{-1})$. Since the excited states
are standing waves, it is easy to see that $c_{mk}^2\sim O(L^{-1})$.
Altogether, we obtain $d\theta/dt\sim O(L^{-2})$.

Actually there is a more elegant way of using adiabatic rotation in
this example. So far we have been taking $\theta$ from 0 to
$\frac{\pi}{4}$. If we instead consider $\theta$ evolving from
$\frac{\pi}{2}$ to $\frac{\pi}{4}$, we would take a single basis
state to a uniform superposition. This suggests that we try: \bea
H(\theta)&=&\cos^2{\theta}\ket{\gamma_0}\bra{\gamma_0}+\sin^2{\theta}\ket{\gamma_{1}}\bra{\gamma_{1}}\nonumber\\
&&-\sin{\theta}\cos{\theta}(\ket{\gamma_0}\bra{\gamma_{1}}+\ket{\gamma_{1}}\bra{\gamma_0})\nonumber\\
&&+\frac{1}{2}\sum_{t=1}^{L-1}\ket{\gamma_t}\bra{\gamma_t}+\ket{\gamma_{t+1}}\bra{\gamma_{t+1}}\nonumber\\
&&-\ket{\gamma_t}\bra{\gamma_{t+1}}-\ket{\gamma_{t+1}}\bra{\gamma_t}
\eea Here $\ket{\gamma_0}$ plays the role of $\ket{\gamma_m}$ in
Eq.\ref{ground}; by rotating $\theta$ from $\frac{\pi}{2}$ to
$\frac{\pi}{4}$, we achieve the desired evolution, remarkably,
leaving most terms constant. The only thing we need to check is the
energy gap at the beginning, where the Hamiltonian is a kinetic
energy term plus a delta function potential of two energy units at
one end. This is still easy to analyze compared to the linear
interpolation. For sufficiently large $L$, the constant potential
just amounts to a zero boundary condition for the low lying
wavefunctions. The lowest energy state in the "box" is a standing
wave with wavelength $4L$, so the gap is also $O(L^{-2})$. If we
wish to implement the mirror evolution described in \cite{c2a} to
get from the history state to $\ket{\gamma_L}$, a similar rotation
can be used on the other end of the "box".

\section{V. Quantum Search and Optimization}

The preparation method above generates the ground state of a
Hamiltonian; as such it is suggestive of a problem-solving
algorithm. Indeed, let us consider an example that can be analyzed
exactly. In \cite{rc} an adiabatic search algorithm was presented
using linear interpolation to an oracle-like problem Hamiltonian. It
turns out that one can formulate an alternative version using
adiabatic rotation. The only extra resource we need is a
non-computational state that the problem Hamiltonian annihilates (we
may, for instance, add a non-computational state to each qubit to
form a qutrit). Call this state $\ket{i}$ and our setup looks as
follows: \bea
H(\theta)&=&(I-\ket{m}\bra{m})+\sin^2{\theta}\ket{\psi_0}\bra{\psi_0}+\cos^2{\theta}\ket{i}\bra{i}
\nonumber\\&&-\sin{\theta}\cos{\theta}(\ket{\psi_0}\bra{i}+\ket{i}\bra{\psi_0})
\label{searchham} \eea where we have followed the notations of
\cite{rc}: $\ket{m}$ is an unknown solution state we want to obtain
and $\ket{\psi_0}$ is a known state that has non-zero overlap with
$\ket{m}$. The identity operator "$I$" only acts on the
computational space and is zero on $\ket{i}$. Denote the overlap by
$a_0\equiv\iprod{\psi_0}{m}$ (i.e. $a_0=\frac{1}{\sqrt{N}}$ in
\cite{rc}). Adiabatically changing $\theta$ from $\frac{\pi}{2}$ to
$0$ evolves the state $\ket{i}$ to $\ket{m}$. The energy gap
$g_1(\theta)=1-\sin{\theta}\sqrt{1-a_0^2}$, which at
$\theta=\frac{\pi}{2}$ is $O(a_0^2)\sim O(N^{-1})$ if
$a_0=\frac{1}{\sqrt{N}}$. This appears different from the case in
\cite{rc} where the minimal gap is $O(N^{-\frac{1}{2}})$. We must
however take into account $\abs{\bra{k}\frac{dH}{d\theta}\ket{0}}$.
The first excited state is proportional to
$a_0\ket{m}+(\sin{\theta}\sqrt{1-a_0^2}-1)\ket{\psi_0}-\cos{\theta}\sqrt{1-a_0^2}\ket{i}$;
explicit calculation yields: \bea
\abs{\bra{1}\frac{dH}{d\theta}\ket{0}}&\leq
&a_0\frac{\abs{\sin{\theta}-\sqrt{1-a_0^2}}}{1-\sin{\theta}\sqrt{1-a_0^2}}
\nonumber\\ &\sim &O(a_0). \eea Now use Eq.\ref{speed} to set the
rotation speed: $d\theta\sim (g_1^2/a_0) dt$. Integrating this from
$\theta=\frac{\pi}{2}$ to $\theta=0$ and setting
$a_0=\frac{1}{\sqrt{N}}$, we obtain the total evolution time $T\sim
(\sqrt{N}-\sqrt{N-1})^{-1}\sim O(\sqrt{N})$ for large $N$. Quadratic
speedup is achieved as promised.

In principle we may replace $I-\ket{m}\bra{m}$ in Eq.\ref{searchham}
by other problem Hamiltonians to find other ground states. This
would work as long as the gap at $\frac{\pi}{2}$ is sufficiently
large. But if we want to solve NP-complete problems as in
\cite{fggs}, notice that $\ket{\psi_0}\bra{\psi_0}$ is generally
nonlocal and does not respect the bit structure \cite{fail}. Suppose
we set $\ket{i}=\ket{222...}$ by using qutrits with an extra state
$\ket{2}$ and assume $\ket{m}=\ket{000...}$ without loss of
generality, it can be proved that no local Hamiltonian can have a
unique ground state of the form \be \frac{1}{\sqrt{1+a_0^2
\tan^2{\theta}}}(\ket{000...}+a_0 \tan{\theta}\ket{222...})
\nonumber \ee by arguments in \cite{c2a}. Since locality is
important to any realistic algorithm, we need a local version of
adiabatic rotation different from (\ref{arot})-(\ref{ground}). For
example, if $\ket{\psi_0}=\bigotimes_n \ket{+}$ where $n$ is the
number of qubits and $\ket{+}=\frac{1}{\sqrt{2}}(\ket{0}+\ket{1})$,
we may want to replace the driving term by $\sum_n
U\ket{+}\bra{+}U^\dag$ where $U$ rotates between $\ket{+}$ and
$\ket{2}$. We may then expect the ground state to contain a piece
proportional to $\bigotimes_n(\ket{+}+\tan{\theta}\ket{2})$. The
appearance of an extra $(3^n-2^n)$ states, however, can
significantly change the spectrum. More work would be required to
find a feasible local model.

\section{VI. Discussion}

The issue of locality arises in the previous section because our
rotation method requires the initial state to be separated from the
computational states. The same requirement applies to the history
state preparation in Section IV, but there locality was preserved by
use of clock qubits (which ensures the orthgonality of the
$\ket{\gamma_t}$s' - we refer the readers to \cite{adkllr} for
details). This suggests that clever use of ancilla would be
important to the design of a local adiabatic rotation algorithm. We
should note that the form of the driving term (\ref{driving}) is
largely chosen for simplicity and can be generalized. The underlying
idea is that when we have an efficient way of adiabatically evolving
from state A to state B, we may combine it with a time-independent
"diffusion" Hamiltonian $H_0$ to attain a more complicated state B'
- a state that overlaps B but we otherwise may not know how to get
to. Such a perspective can be useful for algorithm design.

Results in this article also raise another issue. In almost all our
examples, adiabatic rotations turn out to have the same efficiency
as the linear interpolations - not only up to the same complexity
class, but down to the order of polynomial. One might wonder if
there is some deep connection between different adiabatic paths
under the same set of Hamiltonian interactions. If such connection
exists, might it be possible to prove the efficiency of a linear
interpolation path by proving the efficiency of a long-winded path
that is easier to analyze? At this point this is an entirely open
question.



\begin{thebibliography}{99}
\bibitem{fggs}
E. Farhi, J. Goldstone, S. Gutmann, and M. Sipser,
arXiv:quant-ph/0001106
\bibitem{rc}
J. Roland and N. Cerf, Phys. Rev. A 65, 042308 (2002); See also W.
van Dam, M. Mosca, U. Vazirani, Proc. 42nd Symposium on Foundations
of Computer Science, 279 (2001)
\bibitem{lidar}
A. Hamma and D. Lidar, arXiv:quant-ph/0607145
\bibitem{BR}
S. D. Bartlett, T. Rudolph, Phys. Rev. A, 74, 040302(R) (2006)
\bibitem{adkllr}
D. Aharonov, W. Van Dam, J. Kempe, Z. Landau, and S. Lloyd, O.
Regev, arXiv:quant-ph/0405098
\bibitem{c2a}
M. S. Siu, Phys. Rev. A 71, 062314 (2005)
\bibitem{kitaev}
A. Kitaev, Annals Phys. 303, 2 (2003)
\bibitem{wen}
M. Levin and X.-G. Wen, Phys. Rev. B. 71, 045110 (2005)
\bibitem{dennis}
E. Dennis, A. Kitaev, A. Landahl and J. Preskill, J. Math. Phys. 43,
4452 (2002)
\bibitem{Hastings}
S. Bravyi, M. Hastings, F. Verstraete, Phys. Rev. Lett. 97, 050401
(2006)
\bibitem{ruskai}
P. Deift, M.-B. Ruskai and W. Spitzer, arXiv:quant-ph/0605156
\bibitem{Messiah}
A. Messiah. Quantum Mechanics. John Wiley \& Sons, New York, 1958; A
more rigorous analysis can be found in S. Jansen, M.-B. Ruskai, R.
Seiler, arXiv:quant-ph/0603175
\bibitem{cluster}
R. Raussendorf and H. Briegel. Phys. Rev. Lett. 86, 5188 (2001); M.
Nielsen, arXiv:quant-ph/0504097
\bibitem{gadget}
J. Kempe, A. Kitaev and O. Regev, arXiv:quant-ph/0406180 ; R.
Oliveira and B. Terhal, arXiv:quant-ph/0504050
\bibitem{fail}
E. Farhi, J. Goldstone, S. Gutmann and D. Nagaj,
arXiv:quant-ph/0512159
\end{thebibliography}
\end{document}